\documentclass[aps,prb,twocolumn,superscriptaddress]{revtex4-2}
\usepackage{graphicx}
\usepackage{float}
\usepackage{afterpage}
\usepackage{multirow} 
\usepackage{array}    
\usepackage{amsmath}
\usepackage[T1]{fontenc}
\usepackage{hyperref}

\graphicspath{{figures/}}
\usepackage{color}

\begin{document}

\title{Electronic structure and oxidation states in high-pressure synthesized isostructural CeCN$_5$ and TbCN$_5$}
 
\author{Amanda Ehn}
\email[]{amanda.ehn@liu.se}
\affiliation{Department of Physics, Chemistry, and Biology (IFM), Link\"oping University, SE-58183 Link\"oping, Sweden}
\author{Florian Trybel}
\affiliation{Department of Physics, Chemistry, and Biology (IFM), Link\"oping University, SE-58183 Link\"oping, Sweden}
\author{Talha Bin Masood}
\affiliation{Department of Science and Technology (ITN), Link\"oping University, Norrköping, Sweden}
\author{Leonid~V. Pourovskii}
\affiliation{CPHT, CNRS, École polytechnique, Institut Polytechnique de Paris, 91120 Palaiseau, France}
\affiliation{Collège de France, Université PSL, 11 Place Marcelin Berthelot, 75005 Paris, France}
\author{Igor A. Abrikosov}
\email[]{igor.abrikosov@liu.se}
\affiliation{Department of Physics, Chemistry, and Biology (IFM), Link\"oping University, SE-58183 Link\"oping, Sweden}

\date{\today}

\begin{abstract}
Understanding the behavior of 4\textit{f} electrons in materials containing rare earth elements is one of the fundamental questions within condensed matter physics. In this work the electronic properties of isostructural CeCN$_5$ and TbCN$_5$, both recently synthesized at extreme pressure, are investigated using Density Functional Theory (DFT) calculations. We include the on-site Coulomb repulsion between localized 4$f$ states within the static DFT+U framework; the DFT+U results are cross-checked with DFT+dynamical mean-field theory (DMFT) calculations within the quasi-atomic (Hubbard-I) approximation.
Despite CeCN$_5$ and TbCN$_5$ being isostructural compounds Ce and Tb show different oxidation states, 4+ and 3+ respectively. This leads to distinctly different electronic properties: the former compound is an insulator, while the latter is a metal. An extra electron which is donated by Ce to the polymeric C-N network is distributed across the network. This leads to a modification of the bond length in CeCN$_5$ compared to TbCN$_5$. Still, the polymeric C-N networks can accommodate the different oxidation states in isostructural lanthanide-carbon-nitrogen (LnCN) compounds. Our results underline that LnCN compounds under high pressure offer a unique platform for probing the interplay between 4\textit{f}-electron behavior and structural complexity.

\end{abstract}

\maketitle

\section{Introduction}
Lanthanide compounds attract great attention in condensed matter physics as their partially occupied \textit{f}-shells tend to host strongly correlated electrons. The correlation effects in \textit{f}-electron compounds often lead to highly non-trivial phenomena, such as heavy-fermion behavior \cite{Kotliar} or unconventional superconductivity \cite{Pfleiderer}. Besides being a platform for theoretical and experimental studies of many-electron effects, lanthanide compounds are used in numerous applications, ranging from fuel-cell technologies \cite{Steele2001} to permanent magnets \cite{Sagawa}. The diverse electronic configurations of the partially occupied \textit{f}-shells contribute to the structural complexity of the lanthanide compounds, with trivalent 4\textit{f} metal ions displaying a large variety of coordination numbers, from six to ten \cite{Jones}. This complex interplay between the electronic and the crystal structure in such correlated materials can be particularly well investigated at high pressure conditions, as it allows modification of inter-atomic distances, band widths, hybridization and influences the relative stability between different phases. 

In the broadest terms, 4\textit{f} electrons in such systems may behave either localized or delocalized (itinerant),  which has significant impact on many physical properties of lanthanides; in pure form as well as in compounds. The state of these electrons, \textit{i.e.} whether they are localized or delocalized, may change upon compression \cite{SAMUDRALA2013275}. For example, the isostructural $\alpha\rightarrow\gamma$ transition in Ce discovered by Bridgman at a pressure of $\sim0.8$ GPa almost 100 years ago \cite{Bridgman} is an important example of the influence of pressure on \textit{f}-electrons in lanthanides, that continues to attract significant interest and intense discussions \cite{Amadon2006,Soderlind2025}. Recent advances in high-pressure (HP) synthesis allow not only exploring materials behavior at pressures up to 1 TPa (1000 GPa) \cite{Dubrovinsky2015,Dubrovinsky2022}, but also enable synthesis and detailed experimental investigation of physical and chemical properties of such materials with complex electronic and crystal structures. However, HP synthesis experiments are still mostly focusing on compounds of simple elements, \textit{e.g.}, hydrogen, oxygen, carbon, nitrogen and transition metal elements \cite{DAC_review}. However it was recently shown that it is possible to synthesize binary carbon nitrides at pressures above 100 GPa, which are quenchable to ambient conditions. Being a realization of a theoretical prediction made by Liu and Cohen in 1989 \cite{AmyLiu}, the discovered compounds show calculated hardness values nearing that of diamond, high energy density, piezoelectric, and photoluminescence properties \cite{LanielTrybel2024,C3N4}. Furthermore, the possibility of phonon-induced superconductivity upon hole doping in one of the compounds (\textit{tI}14-C$_3$N$_4$) was recently predicted from theory \cite{Rudenko}. 

\begin{figure*}[ht!]
  \includegraphics[width=\textwidth]{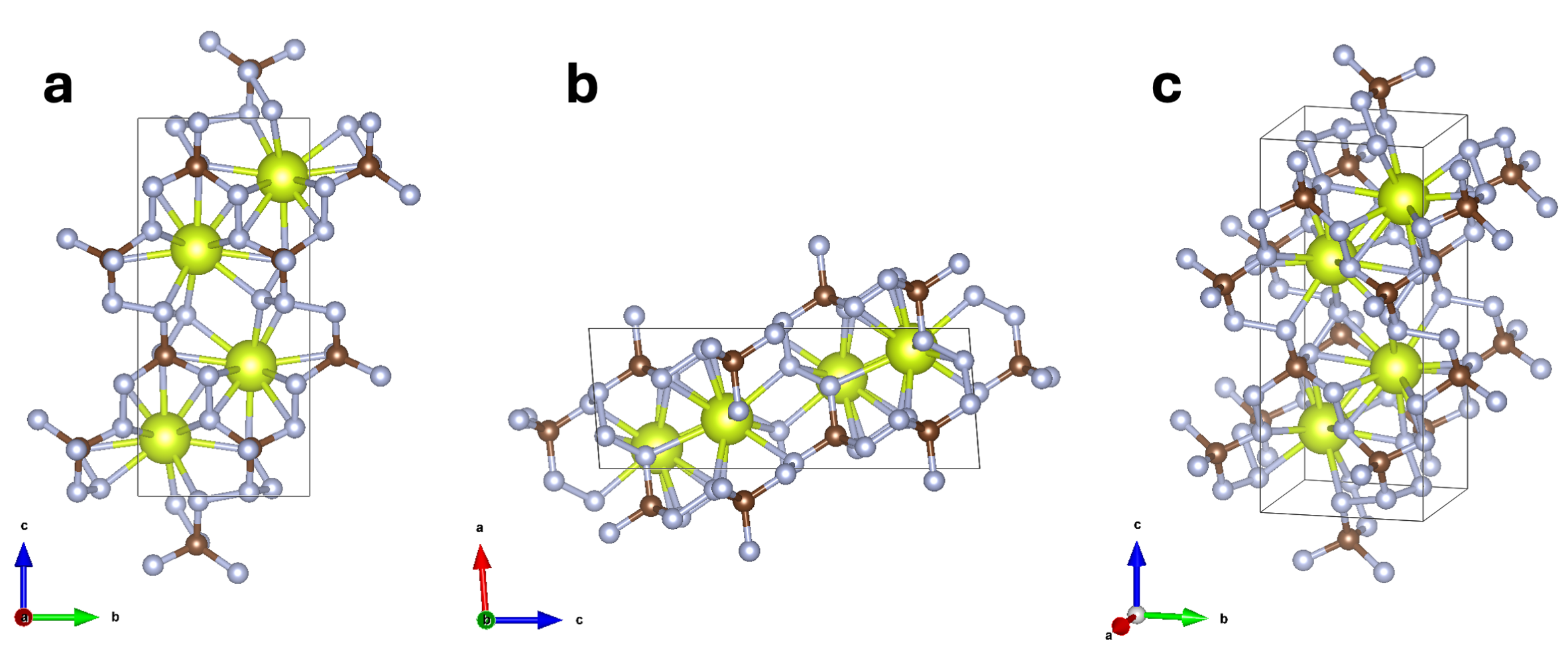}
  \caption{Crystal structure of isostructural LnCN$_5$, here represented by the CeCN$_5$ structure at 90 GPa. Bright green spheres are the lanthanide atom (Ce or Tb), brown spheres C, and silver spheres N.}
  \label{structure}
\end{figure*}

Motivated by the recoverability of the binary compounds,  the experimental studies of carbon nitrides have been extended towards multi-component systems \cite{Bruning}, exploring the possibility of tuning and controlling the physical and chemical properties of this materials family. In particular, the feasibility of a HP synthesis of polycarbonitride compounds containing lanthanide elements has been demonstrated \cite{CeTbsynth, C2N57}. The former work has reported the discovery of four new compounds, LaCN$_3$, TbCN$_3$, CeCN$_5$, and TbCN$_5$, featuring previously unobserved anionic single bonded three-dimensional carbon-nitrogen frameworks consisting of CN$_4$ tetrahedra connected via di- or oligo-nitrogen linkers. Isostructural CeCN$_5$ and TbCN$_5$ appear to be particularly interesting, as both Ce and Tb ions are known to exhibit 3+, as well as 4+ valence in different compounds, but expected to have a similar charge state in isostructural compounds. 

In this work we investigate the electronic properties of these isostructural LnCN$_5$ compounds, with a focus on their electronic structure and bonding at the synthesis pressure using DFT-based calculations with the Hubbard correction included for the rare-earth 4\textit{f} shell within the DFT+U formalism \cite{dftu_dudarev}. The DFT+U results are further validated by comparison with DFT+DMFT \cite{Georges,Anisimov1997_1,Lichtenstein1998,Kotliar2006} calculations, which explicitly capture the local-moment paramagnetic phase of the LnCN$_5$ systems. Unexpectedly, our calculations show that Ce and Tb ions have different valence in the isostructural CeCN$_5$ and TbCN$_5$ compounds, 4+ for Ce and 3+ for Tb. While the oxidation state of Ce agrees with the purely ionic approximation, and the compound is an insulator, there is an apparent contradiction between the expected and observed oxidation state of Tb in TbCN$_5$, and the compound is metallic. Our analysis of the bonding in the two compounds demonstrates that despite the difference in the oxidation states of the rare-earth ions, the polycarbonitride framework is capable of accommodating the difference, while preserving the same crystal structure. 

\section{Crystal chemistry of C\MakeLowercase{e}CN$_5$ and T\MakeLowercase{b}CN$_5$ compounds}

The crystal structure of the isostructural compounds CeCN$_5$ and TbCN$_5$ in monoclinic space group P2$_1$/n (\#14) is shown in Figure \ref{structure} for the example of CeCN$_5$. Both compounds have been previously synthesized in laser-heated diamond anvil cell experiments at a pressure of 90 GPa for CeCN$_5$ and 111 GPa for TbCN$_5$ and their structure was refined through single crystal X-ray diffraction. \cite{CeTbsynth} The crystal structure consists of seven crystallographically unique atoms occupying 4\textit{e} Wyckoff positions: one metal atom (Ce or Tb), one carbon atom, and five nitrogen atoms. There are four formula units in the conventional unit cell (see \textit{e.g.} Fig. \ref{structure}c). Carbon and nitrogen atoms form CN$_4$ tetrahedra linked either by N-N dimers or through trimer bridges with all C and N atoms being \textit{sp}$^3$ hybridized (\textit{c.f.} Fig. 2 of \citet{CeTbsynth}). Based on the bond length distribution, four of the five nitrogen atoms form single bonds, suggesting a charge state of the CN$_5$ unit of 4+ in ionic approximation. Indeed, both Ce and Tb ions are known to exhibit 3+, as well as 4+ valence in different compounds. Thus, Ce, with its ground state electron configuration  ["Xe"]4\textit{f}$^1$ 5\textit{d}$^1$6\textit{s}$^2$ should have 1 occupied \textit{f}-state in the former case and an unoccupied \textit{f}-band in 4+ configurations. Tb (["Xe"]4\textit{f}$^9$ 6\textit{s}$^2$) has 7\textit{f} electrons occupying the spin up states, which are most often well localized and do not contribute to bonding. Therefore, with respect to its chemical bonding in compounds, it should behave similar to Ce: there is either 1 electron occupying the spin down \textit{f}-states in the 3+ configuration, which is the most common for Tb, or its spin-down \textit{f}-states should be empty in a 4+ configuration. Giving this simple consideration, one expects the same valence for Ce and Tb ions in CeCN$_5$ and TbCN$_5$ compounds \cite{CeTbsynth}.

\section{Computational Methodology}
We use the Vienna Ab Initio Package (VASP) \cite{Kresse1, Kresse2, Kresse3, Kresse4} for the DFT+U calculations to properly account for the 4\textit{f} electrons. The DFT+U implementation of \citet{dftu_dudarev} is used, where only the effective U is of importance. We expand further on the choice of U for each compound in the Supplementary Material (SM) \cite{suppl}.

For both compounds the generalized gradient approximation by Perdew-Burke-Enzerhof (PBE) for exchange and correlation was used \cite{PBE}. All simulations used a convergence threshold of 10$^{-7}$ eV. Additionally, for structural relaxations a convergence threshold for forces on individual atoms was set to 10$^{-3}$ eV/Å. The energy cutoff for the plane wave expansion was  set to 680 eV for TbCN$_5$ and 740 eV for CeCN$_5$. A gaussian smearing of 0.03 eV was used for relaxations. The Monkhorst-Pack \cite{MPack} k-point mesh of 6x6x6 was used for relaxations. For calculations of the electronic density of states a mesh of 8x8x8 was used. For this the tetrahedron method with Blöchl corrections without smearing was used. 

For the detailed analysis of the charge distribution, we performed additional calculations with a denser of k-point mesh of 10x10x10. We used a weighted Voronoi diagram-based partitioning approach~\cite{masood2021visual, abrikosov2021topological} to estimate the charge per atom in these two compounds. See SM \cite{suppl} (Sec. S-VII) for detailed description of this method.

In order to cross-check the DFT+U results, we also  carried out  calculations of the two systems in the local-moment  paramagnetic phase using DFT+DMFT and treating local correlations on 4$f$ shells  within the quasi-atomic Hubbard-I (HI) approximation~\cite{Hubbard1963,Lichtenstein1998}; the method is abbreviated below as DFT+HI. 
In these calculations we employed the charge self-consistent DFT+DMFT implementation of Refs.~\cite{Aichhorn2009,Aichhorn2011}, which is based on the full-potential Wien2k DFT code~\cite{Wien2k} and the TRIQS library~\cite{Parcollet2015,Aichhorn2016} implementation of DMFT, together with the Hubbard-I solver provided by the MagInt code~\cite{magint}. Further details of our DFT+HI calculations are given in Section S-I in SM \cite{suppl}.  

\section{Results}
\subsection{Structural Parameters}
The crystal structure for the isostructural LnCN$_5$ compounds is shown in Fig. \ref{structure}. The theoretical structural results are compared to experimental data in Table S-II in SM \cite{suppl}. Starting with CeCN$_5$, the structural parameters are in good agreement with experimental data (see Ref. \cite{CeTbsynth}). Experimental data points are available for pressures as low as 30 GPa, with a full single crystal refinement available at 90 GPa (\textit{i.e.} the structure was previously fully solved from single crystal X-ray diffraction \cite{CeTbsynth}). We find a slightly larger volume compared to the experimental values, about 1.6\% at synthesis pressure. This is in agreement with PBE's general tendency to underbind (see Table S-II in SM \cite{suppl}). 

For TbCN$_5$, only one experimental data point is available; at synthesis pressure for 111 GPa \cite{CeTbsynth}. Here we find nearly perfect agreement with the experimental volume when using PBE+U, with U = 4 eV. The lattice parameters are further well reproduced, with errors within 0.9 -- 3.5\%, where only the $c$-lattice parameter has an error larger than 1 \%. Thus the structural results for both CeCN$_5$ and TbCN$_5$ reproduce the experimental results well.

\subsection{Electronic structure}
\begin{figure}[h]
\centering
\includegraphics[width=0.48\textwidth]{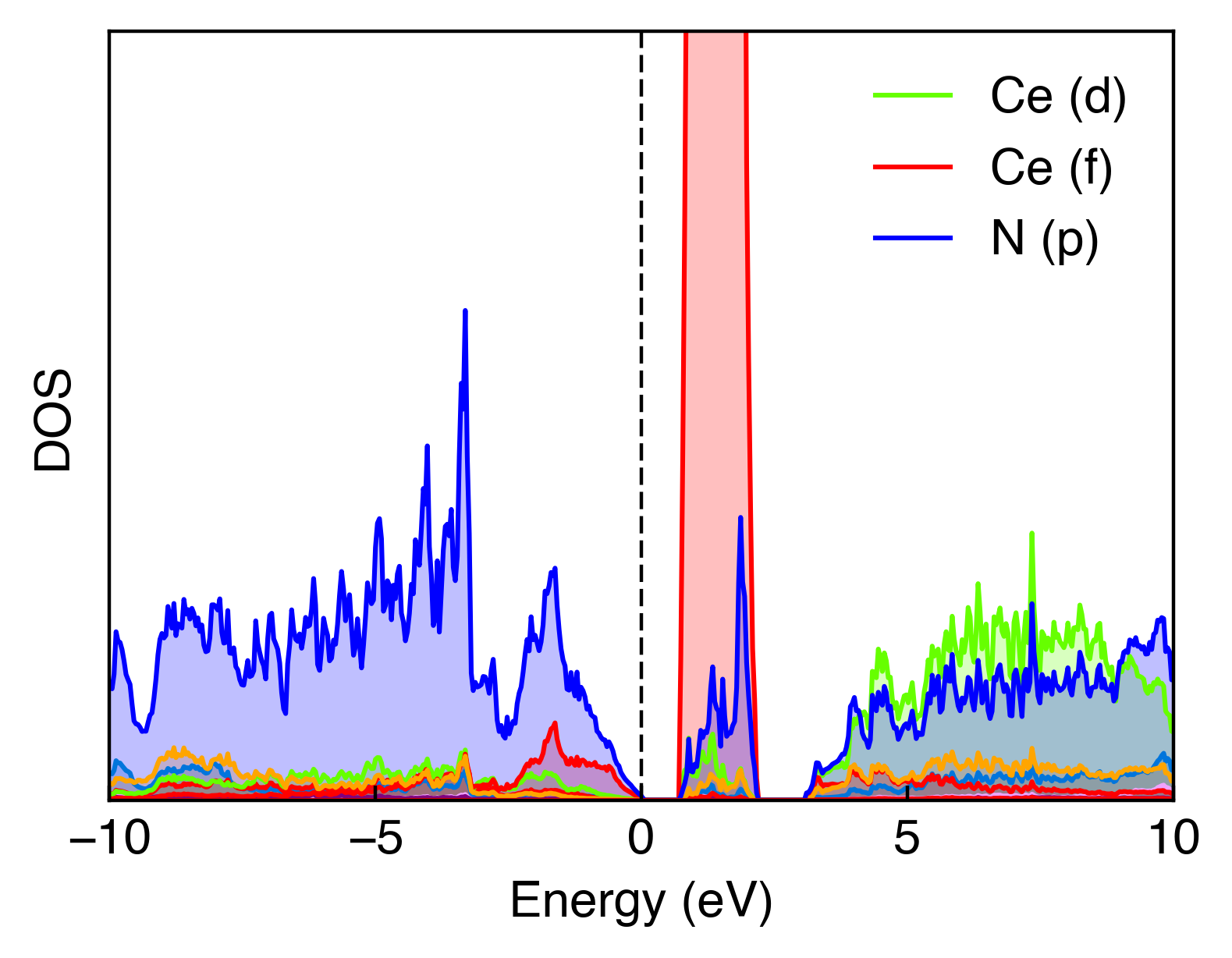}
\includegraphics[width=0.48\textwidth]{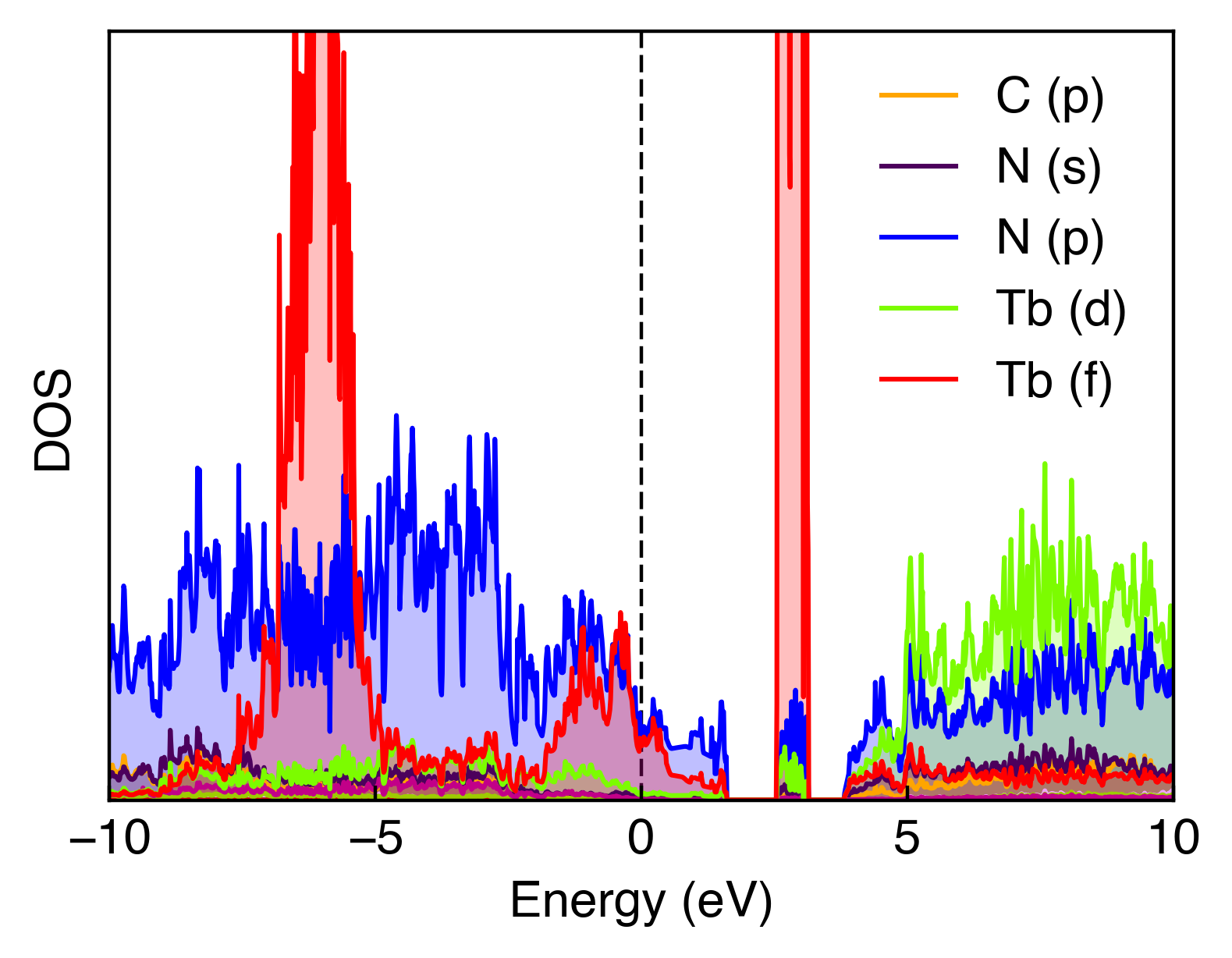}
\caption{Electronic density of states, decomposed onto the orbitals. \textbf{(Top)} CeCN$_5$ at synthesis pressure 90 GPa. \textbf{(Bottom)} TbCN$_5$ at synthesis pressure 110 GPa. Note that the total density of states is not plotted. Fermi energy is adjusted to 0.}
\label{synth_DOS}
\end{figure}

In Figure \ref{synth_DOS} we show the electronic density of states (DOS) for CeCN$_5$ and TbCN$_5$ calculated within the DFT+U theory. Panel (a) shows densities of states projected onto Ce \textit{d}- and \textit{f}-states and onto N \textit{p}-states. Electronic structure calculations show that CeCN$_5$ at its synthesis pressure (90 GPa) is a non-magnetic insulator. The highest occupied valence band is composed of N 2\textit{p} orbitals. The narrow empty band separated by 0.64 eV from the valence band mainly consists of \textit{f} states of Ce. We also see an unoccupied band composed mainly of \textit{d} states of Ce and \textit{p} states of N situated above the empty \textit{f} band of Ce. The bottom of this band is separated from the valence band by 3 eV. 
We note that a small admixture of Ce \textit{f}-states to the valence band and N \textit{p}-states to the Ce-related narrow peak above the Fermi energy, which could be an artifact of the projection \cite{Andersen2000LMTO}.
Indeed, integration of this peak of unoccupied \textit{f}-states gives a value of 14 states per Ce atom, implying that Ce has no occupied \textit{f}-states. Thus, the Ce-atoms are in a 4+ oxidation state, in agreement with the estimate from the previous study \cite{CeTbsynth}. The picture presented for CeCN$_5$ is furthermore in good agreement to what is seen in other Ce$^{4+}$ compounds, like CeO$_2$ \cite{CeO, Baroni}.

\begin{figure}[h]
    \centering
    \includegraphics[width=0.45\textwidth]{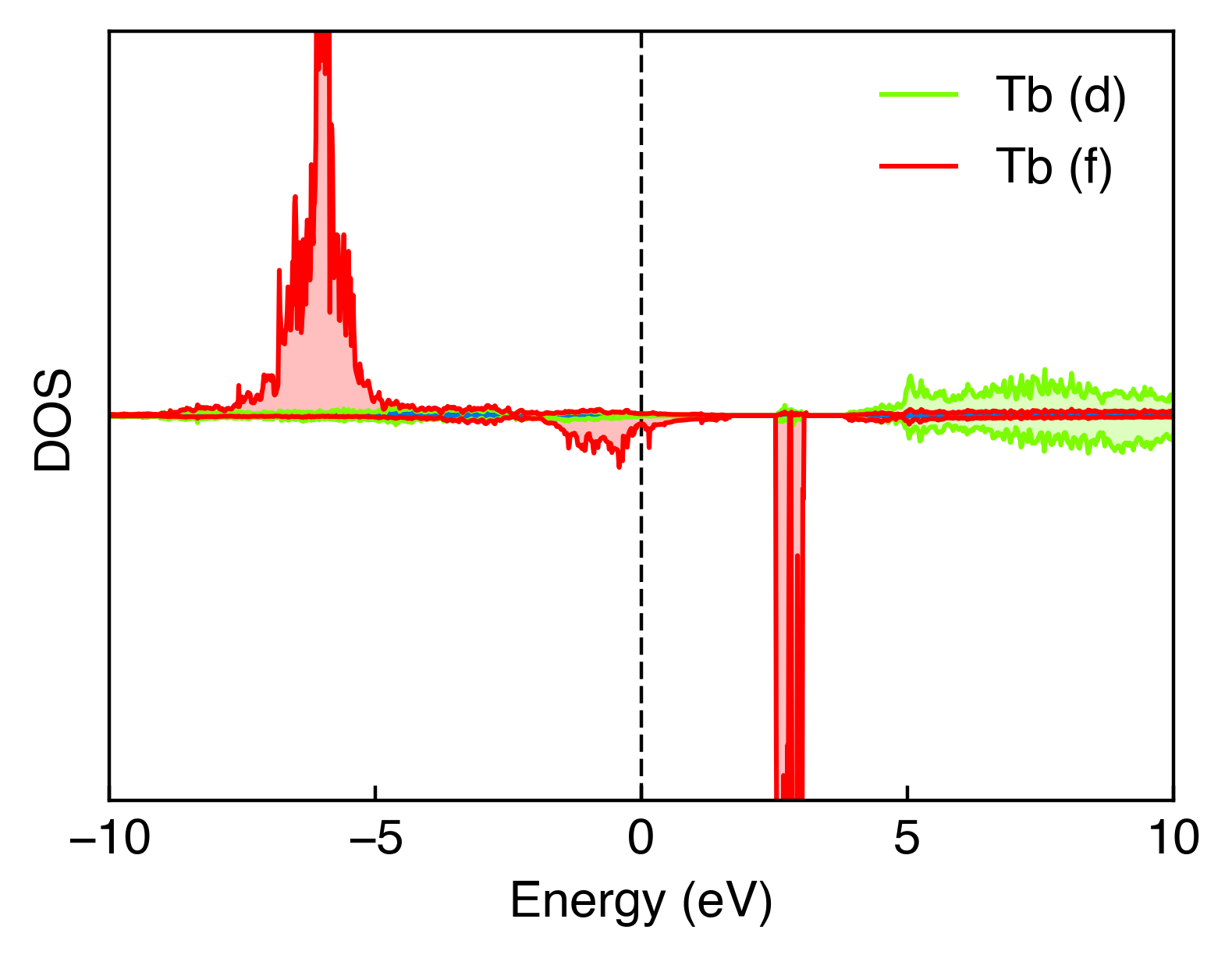}
    \caption{\textbf{(Bottom)} Site-projected density of state for one Tb atom in TbCN$_5$ at synthesis pressure. Fermi energy is adjusted to 0.}
    \label{fig:spin-dos}
\end{figure}

As has been discussed above, the isostructural TbCN$_5$ compound is expected contain Tb-ions in 4+ oxidation state, the same oxidation state as Ce in CeCN$_5$ \cite{CeTbsynth}. Thus, the electronic structure of the two compounds should be similar, and both should be insulators with empty minority spin \textit{f}-states (for Ce the majority spin states are empty as well).  Surprisingly, the electronic DOS of TbCN$_5$ shown in Fig. \ref{synth_DOS}b and \ref{fig:spin-dos} (see also Fig. S4 in SM \cite{suppl}) is qualitatively different from that of CeCN$_5$: according to our calculations TbCN$_5$ is a metal. 

As expected, 7\textit{f} electrons of Tb occupy the majority spin states, which give rise to a narrow peak at about -6 eV below the Fermi energy ($E_\mathrm{F}$). The states are well localized and do not contribute to bonding. Simultaneously, one electron occupies the minority \textit{f}-state of Tb, \textit{i.e.} there is a peak in the electronic DOS projected to the \textit{f}-states of Tb right below $E_{\rm F}$. The occupied orbital is similarly well localized  and is separated from the remaining unoccupied minority spin \textit{f}-states of Tb located at about 2.97 eV above $E_{\rm F}$. Integration of the sharp peak of the total DOS in the energy window corresponding to the unoccupied \textit{f} states of Tb results in 6 empty states per Tb atom. We conclude therefore that the oxidation state of Tb in TbCN$_5$ at the synthesis pressure is 3+. Consequently, Tb donates one less electron to the  [CN$_5$]$_\infty$ anion (${\infty}$ indicates that the CN$_5$ unit is not isolated but embedded in an extended network) compared to Ce,  resulting in a [CN$_5$]$_{\infty}^{3-}$ while the CN$_5$ units bear a charge of -4 in CeCN$_5$ \cite{CeTbsynth}. This is compensated for by the shift of E$_{\rm F}$ into the valence N \textit{p}-states, making the compound metallic at the synthesis pressure (111 GPa). 

Note that TbCN$_5$ is also magnetic, in contrast to CeCN$_5$. The magnetic configuration of the relaxed TbCN$_5$ structure shows strong dependence on the input magnetic moments in the calculations. There are no significant changes in the electronic or structural parameters as a function of the magnetic configurations. The energy differences between magnetic configurations are very small, and a search for the exact magnetic ground state of TbCN$_5$ is beyond the scope of this study. The magnetic configuration for the TbCN$_5$ results presented here can be seen in Fig. S7 in SM \cite{suppl}. 

Furthermore, the electronic structure of both compounds in their paramagnetic phase has been  calculated using DFT+HI. The resulting DFT+HI spectral functions, shown in Figs. S1 and S2 in SM \cite{suppl}, qualitatively agree with DFT+U DOS for the magnetically-ordered case, predicting the Ce$^{4+}$ valency in CeCN$_5$ and the Tb$^{3+}$ one in TbCN$_5$. The multiplet structure of the Tb$^{3+}$ 4$f$ lower and upper Hubbard bands in TbCN$_5$, which is defined by the intra-atomic spin-orbit and Hund's rule interactions, agrees with previous calculations ~\cite{Lebegue2006,Peters2014,Locht2016} and experimental photoemission measurements~\cite{Lang1981} in Tb systems. The alternative Ce$^{3+}$ and Tb$^{4+}$ valencies, which one can try to impose in DFT+HI by the corresponding choice of the double counting correction, are found unstable, see the SM for details \cite{suppl}.

\subsection{Bonding}
\begin{figure}[h]
    \centering
    \includegraphics[width=0.95\linewidth]{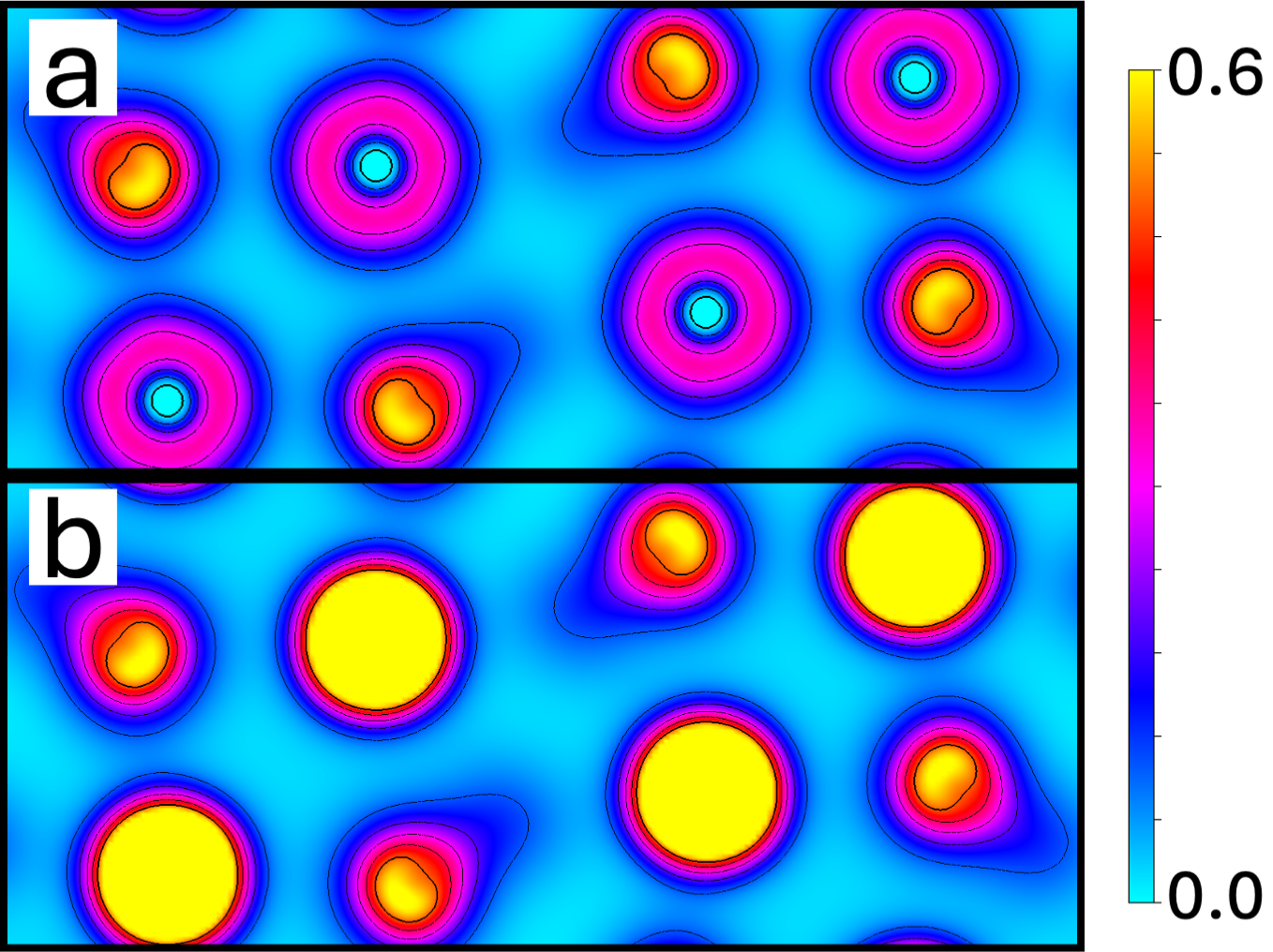}
    \caption{Charge density plot for \textbf{(a)} CeCN$_5$ at 90 GPa, and \textbf{(b)} TbCN$_5$ at 110 GPa. The plane cutting the unit cell is shown in Fig. S5 in SM \cite{suppl}.}
    \label{fig:charge_Ce_Tb}
\end{figure}

The charge density at the respective synthesis pressure for CeCN$_5$ and TbCN$_5$ is shown in  Fig. \ref{fig:charge_Ce_Tb} for a plane cutting diagonally through the unit cell, covering the 4 Ce/Tb atoms, and 4 nitrogen atoms. The  stronger localization in yellow, clearly shows that the additional 4\textit{f}-electron in the Tb structure is localized at the Tb atom. The higher charge at the Ln sites are explained by the charge density being integrated over all valence electrons, and Tb has additional valence electrons compared to Ce. The corresponding electron localization function is presented in Fig. S6 in SM \cite{suppl}, which shows that the electron localization is near identical in both systems despite the difference in oxidation state of the lanthanide atoms.

\begin{figure}[h]
\centering
\includegraphics[width=0.98\linewidth]{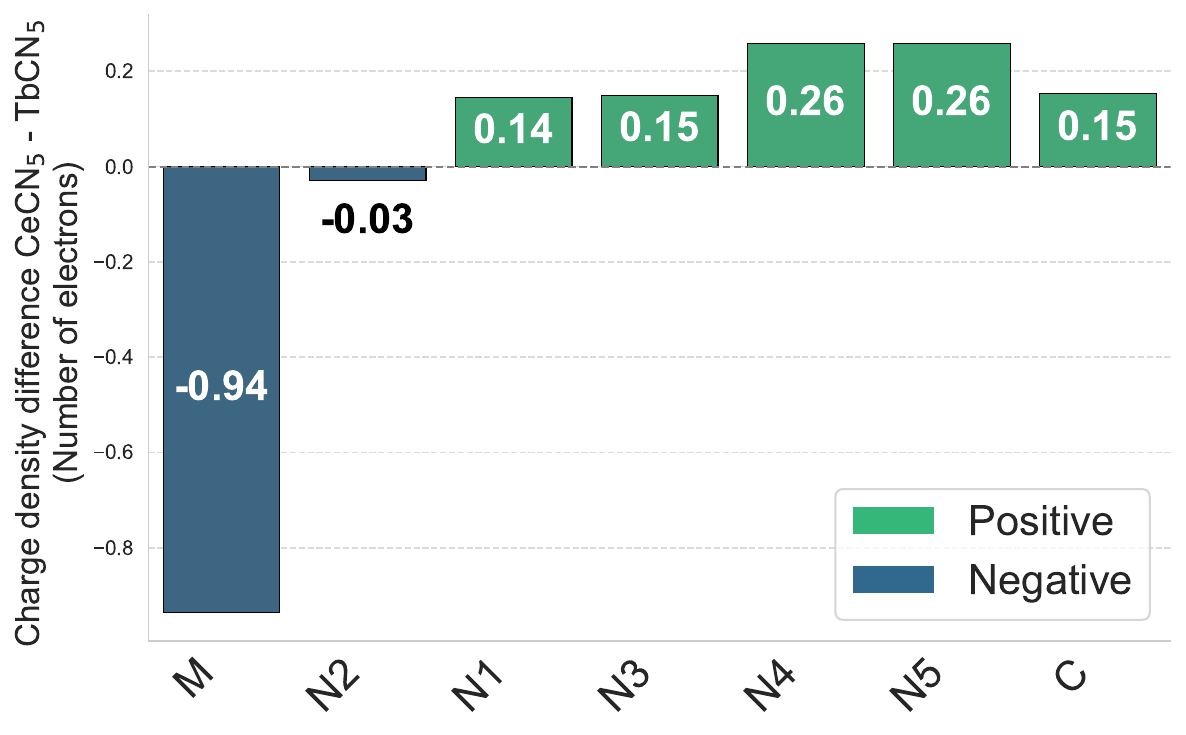}
\caption{Weighted Voronoi charge analysis done on relaxed structures at 90 GPa (CeCN$_5$) and 110 GPa (TbCN$_5$) with ionic radii of 0.97 Å for Ce$^{4+}$ and 1.04 Å for Tb$^{3+}$; \textit{c.f.} Fig. S8 in SM \cite{suppl} for explanation on the labeling of atoms.}
\label{charge:Weighted_voroni}
\end{figure}

To quantify the difference in charges and relate it to the oxidation states of Ce and Tb, a topological analysis of the charge density was performed for the two compounds at their respective synthesis pressure (\textit{c.f} Fig. \ref{charge:Weighted_voroni}). It shows that the two Ln atoms are in different oxidation states, with approximately a value of 0.94 more negative charge localized at the Tb atoms compared to Ce. The additional charge on the Tb atom leads to a charge depletion across the C and N atoms in the C-N network in TbCN$_5$ compared to CeCN$_5$. Alternatively, considering the distribution in CeCN$_5$, the additional electron available to the network due to the Ce$^{4+}$ state, leads to a delocalized distribution across the C-N network and not to a transfer to a single C or N atom (Fig. \ref{charge:Weighted_voroni}).

\begin{figure}[h]
    \centering
    \includegraphics[width=0.9\linewidth]{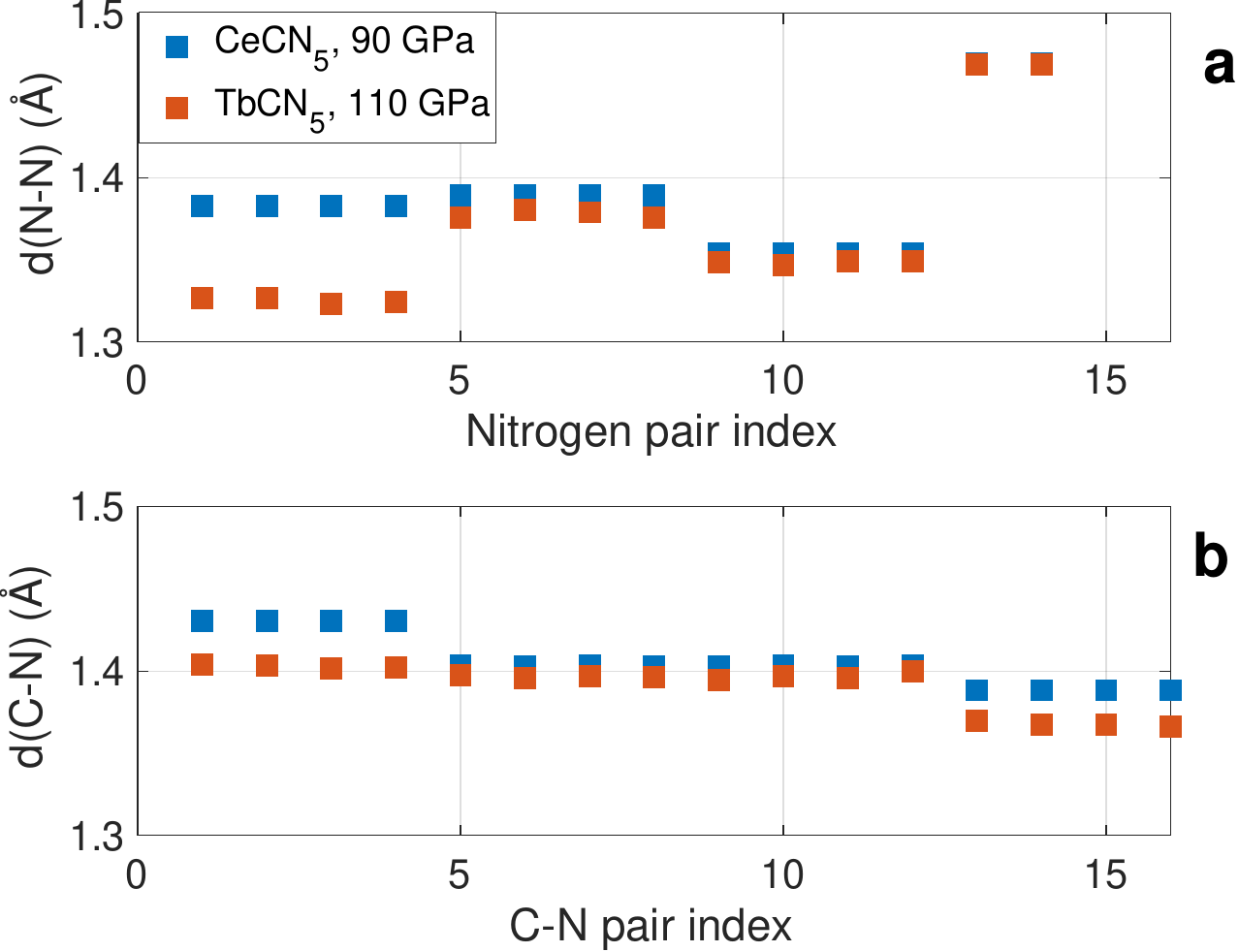}
    \caption{Bond distances in CeCN$_5$ and TbCN$_5$ for (\textbf{a}) N-N distances and (\textbf{b}) C-N distances, at their respective synthesis pressure: 90 and 110 GPa.}
    \label{NN-bonds}
\end{figure}

The extra electron donated from Ce compared to Tb suggest an adaptation of the C-N network; yet the difference in the bonding structure is small. Comparing nitrogen - nitrogen distances and carbon - nitrogen distances shows that there is a contraction in some nitrogen bonds (Fig. \ref{NN-bonds}). The contraction is of the order of 0.05 Å, while the difference in C-N bonds is 0.027 Å. Similar to additional negative charge on molecules leading to an expansion of the bonding length (e.g. O$_2$ vs. O$_2^{-}$), it was shown by \citet{Laniel_Ndimers} that the inter-atomic distance in a nitrogen dimer in nitrides can be related to the formal charge of the dimer unit. A distance difference of 0.060 Å would correspond to a charge difference of half an integer (0.5) formal charge. This suggests that the generally larger N-N distances in CeCN$_5$ compared to TbCN$_5$ are a result of increased negative charge in the C-N network. It needs to be noted that the above mentioned study \cite{Laniel_Ndimers} considered either free units or semi-free $N_2$ dimers, in binary systems where the nitrogen pair connects octahedra. Here we find a qualitatively similar behavior for the nitrogen bonds embedded in the larger C-N network in CeCN$_5$ and TbCN$_5$.


\begin{figure}[h]
    \centering
    \includegraphics[width=0.95\linewidth]{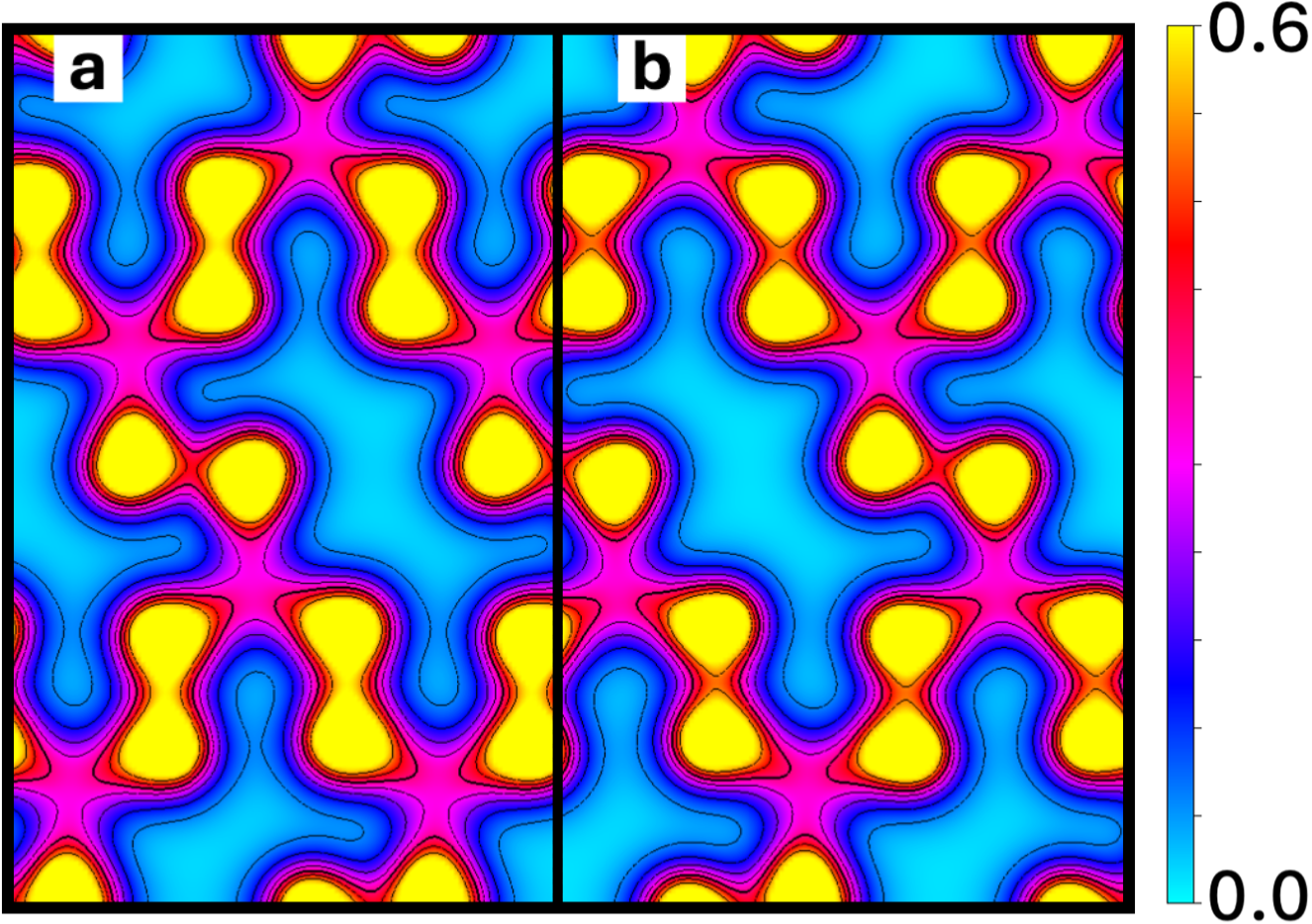}
    \caption{Charge density plot of \textbf{(a)} TbCN$_5$ and \textbf{(b)} CeCN$_5$, with focus on the C-N network. The plane cutting the unit cell is shown in Fig. S5 in SM \cite{suppl}.}
    \label{CHGdensities}
\end{figure}

Lastly the charge densities for CeCN$_5$ and TbCN$_5$ are presented in Fig. \ref{CHGdensities}, cut through a plane covering the C-N network, in particular the nitrogen bonds with the largest difference (as seen in Fig. \ref{NN-bonds}). The charge densities are very similar despite the different oxidation states for Ce and Tb, and the distributed electron over the C-N network. The most notable difference is visible for the vertical nitrogen bonds, which correspond to the nitrogen pair with the largest difference in distance between the CeCN$_5$ and TbCN$_5$ structure. Thus, the shorter nitrogen pair distance in the TbCN$_5$ structure leads to a visually stronger localization of charge, but as seen from the topological charge analysis, the overall integrated charge density is higher in the C-N network in the CeCN$_5$ structure.

\section{Conclusions}
In conclusion, the oxidation states of high-pressure synthesized isostructural CeCN$_5$ and TbCN$_5$ have been investigated. Ab initio  calculations show that the two isostructural compounds are in different oxidation states, with Ce being in 4+ state and Tb in 3+ state. This results in distinctly different electronic structures of the compounds, as the additional electron donated to the C-N network in CeCN$_5$ leads to an insulating state, while the TbCN$_5$ compound is metallic. These conclusions are obtained by DFT+U calculations for magnetically ordered phases and they are further confirmed by direct calculations of the local-moment paramagnetic phase by DFT+DMFT approach within the Hubbard-I approximations. A topological charge analysis applied to the two systems shows that the additional charge is distributed over the complete C-N network, with a slightly dominating localization at N-N bonds. The additional charge induces a small increase in bond distances, indicating that polymeric C-N networks can accommodate different charge states via small adaptations of interatomic distances.  Thus, the results demonstrate that polymeric C–N networks enable the formation of isostructural compounds containing Ln atoms in different oxidation states. This significantly broadens the accessible electronic structure space and may through alloying open up entirely new paths for tuning the electronic properties of this emerging class of materials.

\section*{Acknowledgments}
The authors thank Dominique Laniel (Centre for Science at Extreme Conditions and School of Physics and Astronomy,
University of Edinburgh) for valuable discussions.
A.E. and I.A.A. acknowledge support by the Knut and Alice Wallenberg Foundation (Wallenberg Scholar grant no. KAW-2023.0309) and by the Swedish Research Council (VR) grant no. 2023-05358. Support by the Swedish Government Strategic Research Area (SRA) the Swedish e-science Center (SeRC) and SRA in Materials Science on Functional Materials at Linköping University (Faculty Grant SFO-Mat-LiU no. 2009 00971) is gratefully acknowledged. 

L.V.P. is grateful to the CPHT computer team for support.

T.B.M. acknowledges partial support from Swedish Research Council (VR) grant 2023-04806; Swedish e-Science Research Center (SeRC); and, Wallenberg AI, Autonomous Systems and Software Program (WASP) funded by the Knut and Alice Wallenberg Foundation.

F.T. acknowledges support through ERC Grant (UNMASCC-HP,  10.3030/101117758). Funded by the European Union. Views and opinions expressed are however those of the author(s) only and do not necessarily reflect those of the European Union or the European Research Council Executive Agency. Neither the European Union nor the granting authority can be held responsible for them. 

Computations were enabled by resources provided by the National Academic Infrastructure for Supercomputing in Sweden (NAISS), partially funded by the Swedish Research Council through grant agreement no. 2022-06725. We also acknowledge NAISS for providing access to the LUMI supercomputer, owned by the EuroHPC Joint Undertaking and hosted by CSC (Finland) and the LUMI
consortium.

\bibliography{Ce_Tb_Manuscript.bib}

\end{document}


\title{SUPPLEMENTARY MATERIAL \\ Electronic structure and oxidation states in high-pressure synthesized isostructural CeCN$_5$ and TbCN$_5$}
 
\author{Amanda Ehn}
\affiliation{Department of Physics, Chemistry, and Biology (IFM), Link\"oping University, SE-58183 Link\"oping, Sweden}
\author{Florian Trybel}
\affiliation{Department of Physics, Chemistry, and Biology (IFM), Link\"oping University, SE-58183 Link\"oping, Sweden}
\author{Talha Bin Masood}
\affiliation{Department of Science and Technology (ITN), Linköping University, Norrköping, Sweden}
\author{Leonid Pourovskii}
\affiliation{CPHT, CNRS, École polytechnique, Institut Polytechnique de Paris, 91120 Palaiseau, France}
\affiliation{Collège de France, Université PSL, 11 Place Marcelin Berthelot, 75005 Paris, France}
\author{Igor A. Abrikosov}
\affiliation{Department of Physics, Chemistry, and Biology (IFM), Link\"oping University, SE-58183 Link\"oping, Sweden}

\date{\today}

\maketitle

\section{DFT+HI calculations}\label{sec:dft_hi}

We calculated the electronic structure of paramagnetic CeCN$_5$ and TbCN$_5$ using  DFT+DMFT approach in the quasi-atomic Hubbard-I (HI) approximation (DFT+HI). 
In our DFT+HI calculations we employed the charge self-consistent DFT+DMFT implementation of Refs.~\cite{Aichhorn2009,Aichhorn2011}, which is based on the full-potential linearized augmented-plane-wave (LAPW) Wien2k DFT code~\cite{Wien2k} and the TRIQS library~\cite{Parcollet2015,Aichhorn2016} implementation of DMFT, together with the Hubbard-I solver provided by the MagInt code~\cite{magint}. Projective Wannier orbitals representing 4$f$ states were constructed in accordance with Ref.~\cite{Aichhorn2009} using the Kohn-Sham (KS) bands within the energy window [-0.15:0.32]([-0.15:0.15])~Ry for the Ce (Tb) system, respectively. The energy window is chosen to enclose the 4$f$ KS bands. Its larger upper bound in the case of CeCN$_5$ is due to the empty Ce-4$f$ KS band located significantly above the KS Fermi level; the upper bound is thus set to be above the 4$f$-band upper edge. We employed 200 $\vk$-points in the full Brillouin zone, the LAPW basis cutoff $R_{\mathrm{mt}}K_{\mathrm{MAX}}=7$, and the LDA exchange correlation potential. The spin-orbit was included using the standard second-variation approach.

The rotationally-invariant Coulomb interaction was specified by $U=F^0=6(7)$~eV and the Hund's rule coupling $J_H=0.7 (0.95)$~eV for the Ce (Tb) system, respectively. 
 The values of $U$ and $J_H$ together with the standard assumptions on the ratios of the Slater parameters ($F^4/F^2$=0.668, $F^6/F^2$=0.45 \cite{Freeman1965}) fully determine the on-site Coulomb interaction. 
The chosen $U$ and $J_H$ values are in the standard range employed in the literature for Ce and Tb systems, see Suppl.~Sec.~\ref{sec:U} for further discussion. 

We employ the DMFT double-counting (DC) correction in the fully-localized limit using the nominal integer occupancies for the 4$f$ shell, as shown to be appropriate for DFT+HI~\cite{Pourovskii2007}. The nominal occupancy in DC was set to correspond to either 3+ (Ce f$^1$ and Tb $f^8$) or 4+ (Ce f$^0$ and Tb $f^7$) valency of the corresponding RE ion. 

\section{Paramagnetic electronic structure within DFT+HI}

The DFT+HI calculations were first run to self-consistency to determine the stability of a given 4$f$ valency. The Ce $4f^0$ and Tb $4f^8$ were found to be stable, i.e. self-consistent DFT+HI calculations with the DC term set for a given 4$f$ occupancy converged to the same occupancy. In contrast, with the DC term set for Ce $4f^1$, the calculations still converged to the Ce $4f^0$ occupancy. In the case of TbCN$_5$ run with the DC value corresponding to $4f^7$, the Tb occupancy was fluctuating over iterations between 7 and 8 with no self-consistency reached. Correspondingly, we conclude that the Ce $4f^1$ and Tb $4f^7$ occupancies are unstable.

\begin{figure}[H]
    \centering
    \includegraphics[width=0.5\linewidth]{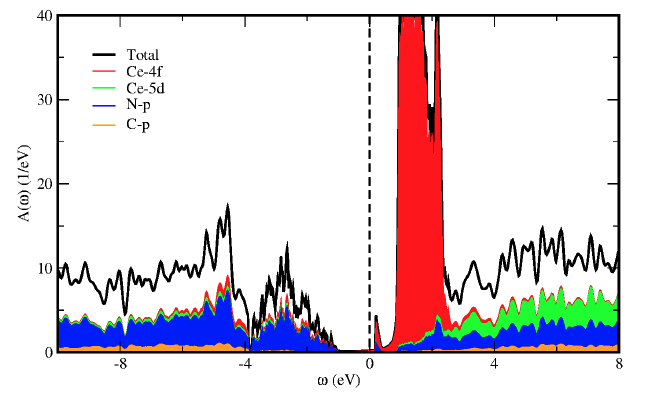}
    \caption{Spectral function of CeCN$_5$ for the 4+ Ce valency calculated with DFT+HI.}
    \label{fig:DMFT_Ce}
\end{figure}

\begin{figure}[H]
    \centering
    \includegraphics[width=0.5\linewidth]{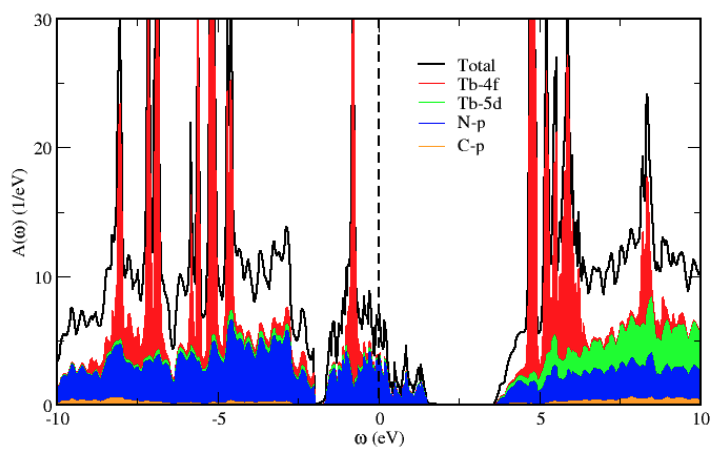}
    \caption{Spectral function of TbCN$_5$ for the 3+ Tb valency calculated with DFT+HI.}
    \label{fig:DMFT_Tb}
\end{figure}

The resulting DFT+HI spectral function for the stable RE valency is displayed in Fig.~\ref{fig:DMFT_Ce} (Fig.~\ref{fig:DMFT_Tb}) for CeCN$_5$ (TbCN$_5$). In the Ce case, the compound is insulating with empty 4$f$ states located at the bottom of the conduction band. The CeCN$_5$ DFT+HI spectral function closely agrees with its DFT+U DOS (Fig.~2 of the main text). In the case of TbCN$_5$, DFT+HI calculations reproduce the qualitative DFT+U picture (Fig.~2 of the main text) of this system being a metal with the Fermi level located within the N-2p band. In addition, DFT+HI  captures multiplet effects in the lower and upper Hubbard bands, i.~e. their splitting into distinct peaks corresponding to various excited configurations of the Tb 4$f$ shell. The structure of those peaks in the case of Tb 3+ has been observed experimentally~\cite{Lang1981} and analyzed in details elsewhere~\cite{Lebegue2006,Lang1981}. For example, the lower-Hubbard-band excitation at about -1 eV corresponds to the Hund's rule maximum-spin 4$f^7$ $^8S_{7/2}$ excited state; this half-filled Hund's rule state is much lower in energy than other 4$f^7$ excited states that are located in the range from -5 to -9 eV. 

\section{Choice of $U$ in GGA+U calculations} 
\label{sec:U}
Choosing the value of the Hubbard parameter $U$ for rare-earth ions in CeCN$_5$ and TbCN$_5$ compounds for our DFT+U calculations is not a straightforward task. Experimentally the value of $U$ can be most directly assessed by simultaneously probing occupied and empty 4$f$ states with  photoemission and inverse photoemission spectroscopies
. The outcome is then compared the with ab initio DFT+U or DFT+DMFT calculations. Such measurements were carried out for rare-earth elemental metals~\cite{Lang1981} and the resulting spectra were well accounted for  by employing $U$=7~eV in DFT+HI~\cite{Locht2016}. A good agreement with experimental spectroscopy for Ce metal is obtain by using $U$ of about 6~eV in full DFT+DMFT and DFT+HI~\cite{McMahan2003,Haule2005,Amadon2006,Pourovskii2007,Huang2019}. Various DFT+DMFT and DFT+HI electronic-structure calculations of rare-earth semiconductors (nitrides, oxides, sulfides) also employ $U$ in the range of 6 to 8 eV~\cite{Pourovskii2007,Amadon2012,Peters2014,Herper2020,Galler2021,Boust2022,Amidani2025}. The similar range is  predicted by recent direct U calculations using constrained-DFT(+U/HI) approaches~\cite{Galler2022,Liu2023}, with a moderate increase in U along the rare-earth  series.  The Hund's rule coupling $J_H$ of rare-earth ions doped into wide-band insulators can be directly measured  in solid-state environment using 
$f-f$ optical transitions~\cite{Carnall1989}. As a result, the $J_H$ value in rare-earth ions  was found to be independent of crystalline environment, with the characteristic spectra of sharp optical excitations for each rare-earth ion~\cite{Carnall1989,Liu2005,Reid2016} reproduced in all insulating hosts. 

Correspondingly, in our DFT+HI calculations we employ the $U$ values of 6(7)~eV for the Ce (Tb) system together with  $J_H$ values specified in Suppl.~Sec.~\ref{sec:dft_hi}  extracted from optical measurements~\cite{Carnall1989}. 

However, the optimal value of $U$ for calculating total energy and ground-state properties in the DFT+U framework can differ from the one giving the best account for spectral properties.  With the DC correction not being exact, the DFT+U framework is sensitive to the underlying approximation for the DFT exchange and correlation functional. In particular, since GGA exhibits a weaker overbinding tendency compared to LDA, the optimal value of $U$ to capture the ground-state volume and equation of state is expected to be smaller in GGA+U as compared to LDA+U.  Typically, when LDA is employed for Ce metal a value of U $\sim$6 eV is used, similar to DFT+DMFT calculations as discussed above, and the same value is often used in its compounds. The value of U is typically lower when GGA+U approach is used, e.~g. for Ce and its oxides \cite{Baroni,CeO}. For terbium compounds values of U ranging from 3-9 eV have been successfully used \cite{Peters2014, RareN, TbNTbP}, depending on choice of the DFT functional (i.e. LDA vs GGA).  While comparing those values with DFT+DMFT(HI) calculations of spectroscopies cited earlier, it is also worth to notice that most of DFT+U works employ the Dudarev formulation   \cite{dftu_dudarev}, which uses only  an effective value of $U_{eff}= U - J_H$. This $U_{eff}$ value is typically cited in the DFT+U literature.

\begin{figure}[H]
\includegraphics[width=0.34\linewidth]{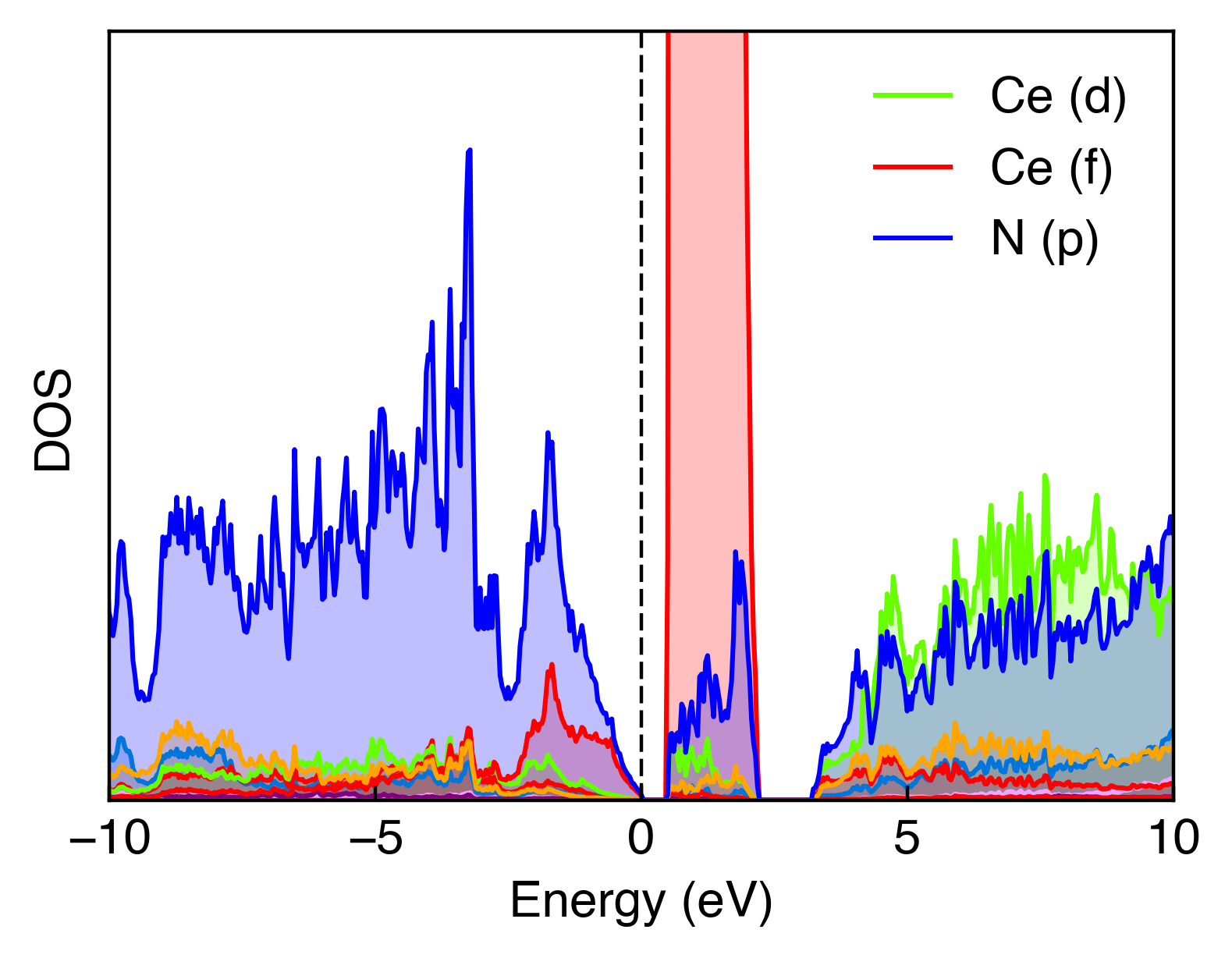}
\includegraphics[width=0.34\linewidth]{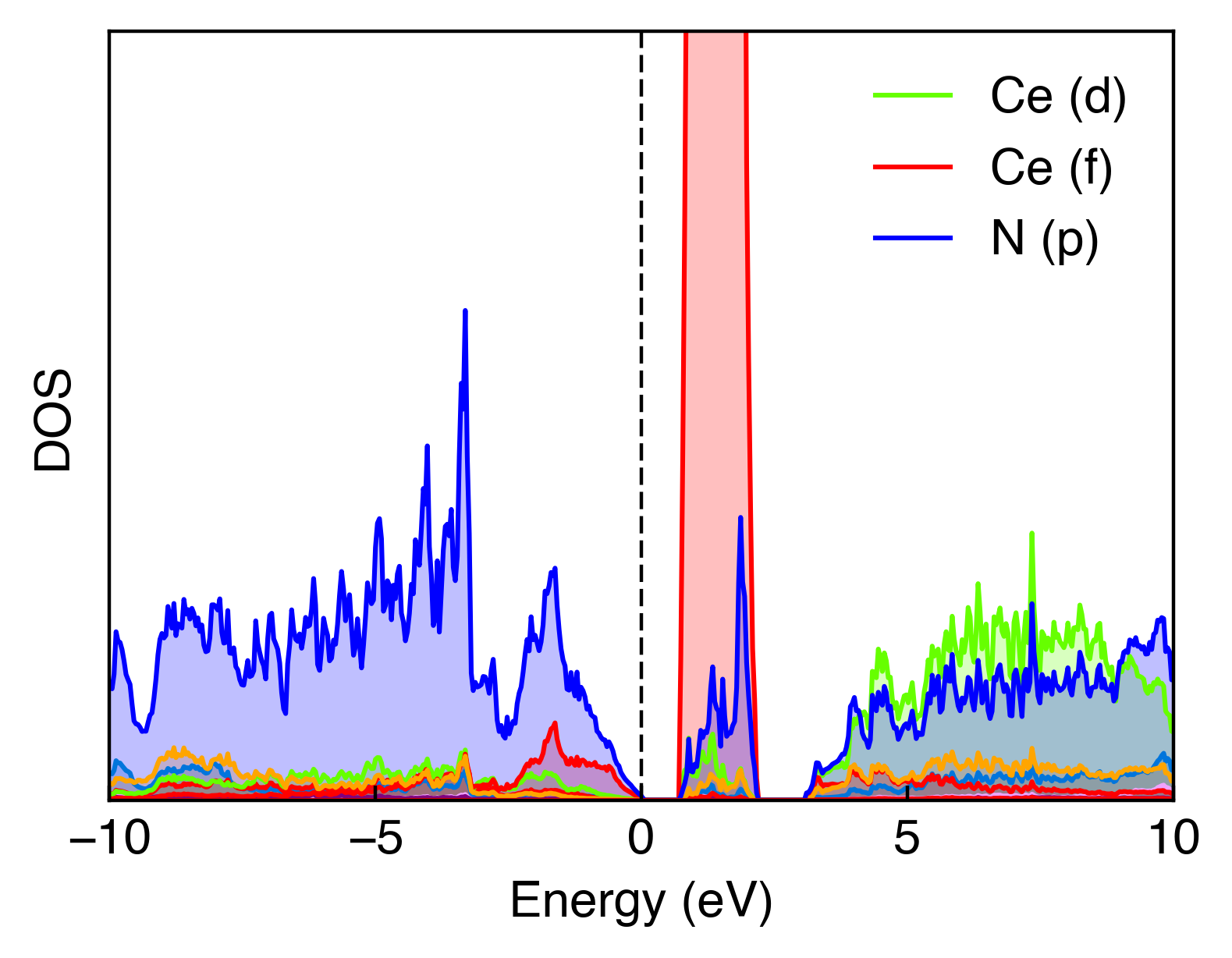}
\includegraphics[width=0.34\linewidth]{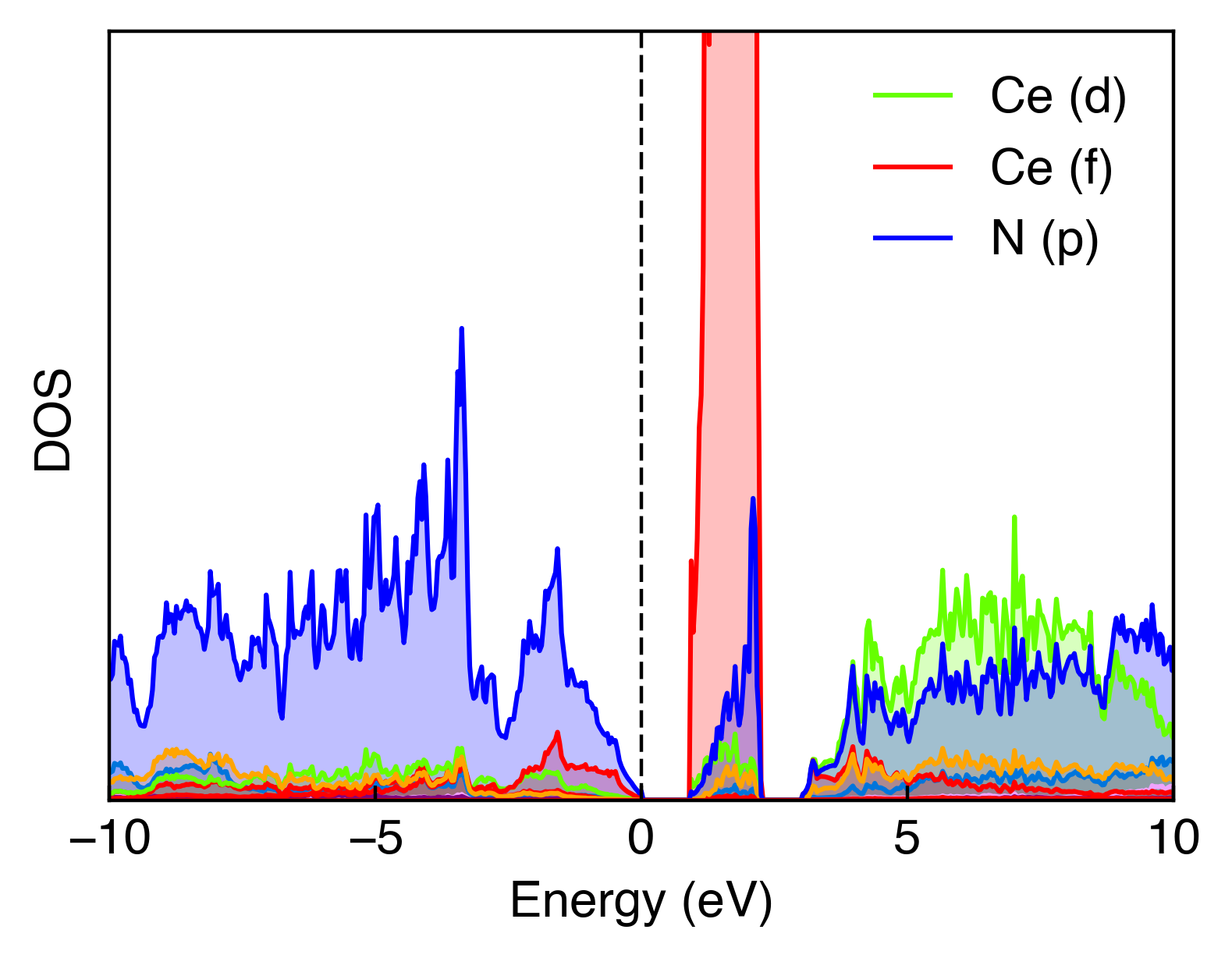}
\caption{Electronic structure of CeCN$_5$ calculated within (a) GGA method (U$_{eff}$ = 0 eV) and  GGA+U method with (b) U$_{eff}$ = 3 eV and (c) U$_{eff}$ = 6 eV at volumes corresponding to 90 GPa with each setting, 194.4 Å$^3$, 195.8 Å$^3$, and 197.2 Å$^3$ for U$_{eff}$ = 0, 3, and 6 eV, respectively.}
\label{comp_Ce_LDAPBE}
\end{figure}

Correspondingly, in order to choose $U_{eff}$ for our GGA+U calculations we first checked its influence  on the electronic structure of CeCN$_5$ and TbCN$_5$ compounds. 
For the CeCN$_5$ compound, we have varied U$_{eff}$ in our GGA+U calculations in the range from 0 to 6 eV, and observed weak dependence of the calculated electronic density of states (DOS) on U$_{eff}$  (Fig. \ref{comp_Ce_LDAPBE}). The increasing value of U slightly shifts the unoccupied \textit{f}-states away from the Fermi energy. At the same time, even with the value of U$_{eff}$ = 0 the \textit{f}-states remain unoccupied, which explains the very small influence of U$_{eff}$ on the DOS. In all cases Ce ion remains in the 4+ configuration. 

We note that the electronic density of states presented in Fig. \ref{comp_Ce_LDAPBE} agrees well with the result obtained for CeCN$_5$ in \cite{CeTbsynth} within LDA+U framework with U$_{eff}$ = 6 eV (see Fig. 5 in Ref. \cite{CeTbsynth}). Importantly, our DFT+U results agree well with the electronic structure of CeCN$_5$ compound calculated in the framework of DFT+DMFT method (Fig. \ref{fig:DMFT_Ce}). 

\begin{figure}[H]
    \centering
    \includegraphics[width=0.5\linewidth]{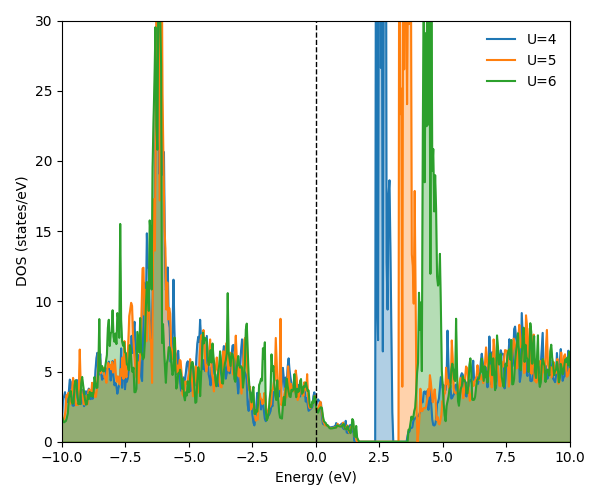}\includegraphics[width=0.5\linewidth]{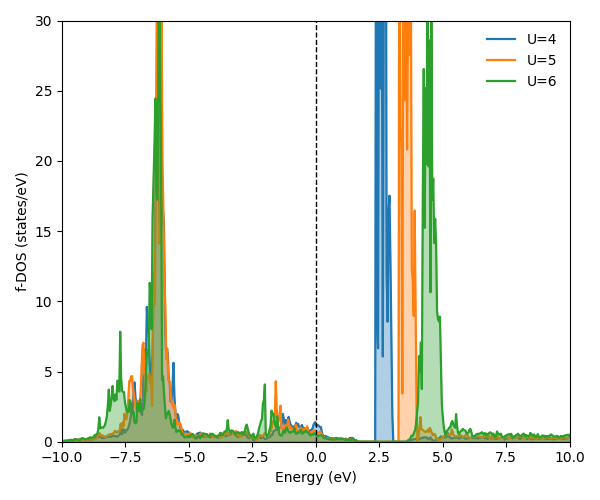}
    \caption{Electronic structure of TbCN$_5$ calculated within GGA+U method with U$_{eff}$ = 4, 5 and 6 eV at experimental volume 178.84 Å$^3$, corresponding (in GGA+U) to pressure 108 GPa, 106 GPa, and 100 GPa for each U$_{eff}$, respectively. The left figure is the total density of states, and the right are the projected density of states of the \textit{f}-states in Tb.}
    \label{fig:comp_Tb_U}
\end{figure}

For TbCN$_5$ compound, we have varied U$_{eff}$ in our GGA+U calculations in the range from 4 to 6 eV. The dependence of the calculated electronic structure on U$_{eff}$  (Fig. \ref{fig:comp_Tb_U}) is stronger than for CeCN$_5$ compound. However, independently of the value of U$_{eff}$ the compound is metallic and there is a peak in the electronic DOS projected to the \textit{f}-states of Tb right below E$_\mathrm{F}$. Thus, our conclusion that Tb ion is in the 3+ configuration does not depend on the U$_{eff}$. 

Finally, we have checked the influence of U$_{eff}$ in our GGA+U calculations on the crystal structure parameters of CeCN$_5$ and TbCN$_5$. The results are summarized in Table \ref{tab:GGAU}. One can see that changing the value of U shifts the volume slightly. For TbCN$_5$ compounds, U$_{eff}$ = 4 eV shows excellent agreement with experimental data regarding the structure parameters. For CeCN$_5$ compounds, even PBE calculations (U$_{eff}$ = 0) lead to a slightly overestimated volume at the synthesis pressure, and the volume increases further with increasing U$_{eff}$. However, the results are still in very good agreement with experimental data measured at the synthesis pressure. We also note that, generally pure GGA calculations tend to overestimate the volume while LDA underestimates the volume. In the DFT+U formalism, it is expected that for the equal U$_{eff}$, the GGA+U method results in a larger volume than the LDA+U method (used in \cite{CeTbsynth})  for the same pressure. 

Considering all the results and discussion presented in this section, we have chosen to show and discuss results calculated within GGA+U with U$_{eff}$ = 3 eV for CeCN$_5$ compound and U$_{eff}$ = 4 eV for TbCN$_5$ throughout the main text of the paper. 

\begin{table}[H]
    \centering
    \caption{Structural parameter dependence of U in GGA+U calculations for CeCN$_5$ and TbCN$_5$.}
    \begin{tabular}{c|ccccccc}
    \hline \hline
       Compound  & U value & V & a & b & c &$\alpha = \gamma$ & $\beta$ \\
    
        & (eV) & Å$^3$ & Å & Å & Å & $^\circ$ & $^\circ$\\ \hline \hline
        CeCN$_5$ & 0& 194.40 &  3.91 & 4.73 &10.95&  90 &106 \\ 
        CeCN$_5$ & 3 & 195.84 & 3.91& 4.75 & 10.89 & 90 & 106 \\ 
        CeCN$_5$ & 6 & 197.17& 3.92  &4.77 &10.99  &90 &106 \\ 
        TbCN$_5$ & 4 & 178.83 & 3.87 & 4.49 & 10.74& 90 & 106 \\ 
        TbCN$_5$ & 5&  177.97 &3.86 & 4.48 &10.72  &90& 106  \\ 
        TbCN$_5$ & 6 & 175.39&  3.84  &4.45 &10.66 & 90 & 106\\ \hline \hline
        \end{tabular}
    \label{tab:GGAU}
\end{table}

\section{Comparison to experimental results of the structural parameters}

\begin{table}[H]
\centering
\caption{Comparison of structural parameters of LnCN$_5$ compounds, at their respective synthesis pressure. Experimental data is taken from Ref. \cite{CeTbsynth}. LnCN$_5$ were calculated with GGA+U where U$_{Ce}$ = 3 eV and U$_{Tb}$ = 4 eV.}
\begin{tabular}{ccccccccccccc}
\hline \hline
\multirow{2}{*}{Compound} & P$_{synth}$ & P$_{DFT}$ & $V_{exp}$ & $V_{DFT}$ & $a_{exp}$& $a_{DFT}$ & $b_{exp}$ & $b_{DFT}$ & $c_{exp}$ & c$_{DFT}$ & $\alpha = \gamma$ & $\beta$\\ 
 & (GPa) &  (GPa) & (Å$^3$) & (Å$^3$) & Å & Å & Å  & Å  & Å  & Å & $\circ$ & $\circ$\\ \hline \hline
CeCN$_5$ & 90  & 90 & 192.72  & 195.84 & 3.89 & 3.91 &4.74 & 4.75306 &10.89440 &10.97138 & 90 & 106 \\ \hline
CeCN$_5$ & 69  & 70 & 198.54  & 203.64 &  &  &  &  & & & & \\ \hline
CeCN$_5$ & 58  & 60 & 204.3  & 208.10  &  &  &  &  & & & &\\ \hline
TbCN$_5$ & 111 & 110 & 178.84  & 178.83 &3.83 & 3.87 & 4.5221&4.48946 & 10.74010 &10.71809 & 90 & 106 \\ \hline
\label{struc_tab}
\end{tabular}
\end{table}

\section{Electron localization function}
\begin{figure}[H]
    \centering
    \includegraphics[width=0.3\linewidth]{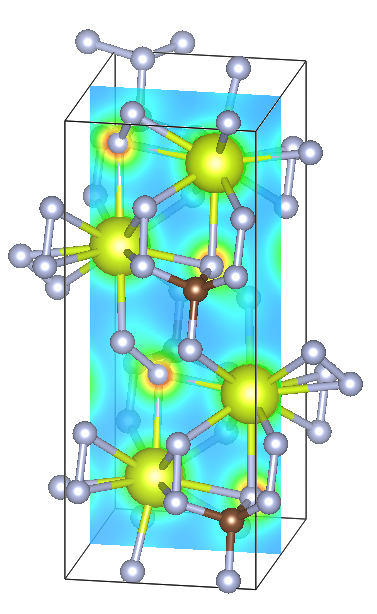}\includegraphics[width=0.3\linewidth]{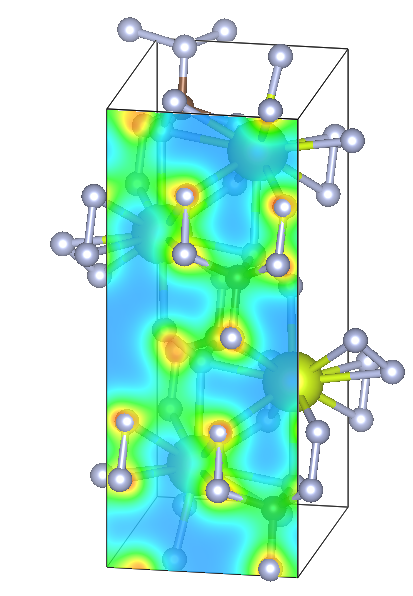}
    \caption{The planes cutting through the unit cell to produce the charge density plots and electron localization plots in Figs. 4 and 7 in the main paper, and Fig. \ref{fig:elf_CeTb}.}
    \label{fig:planes}
\end{figure}

\begin{figure}[H]
    \centering
    \includegraphics[width=0.5\linewidth]{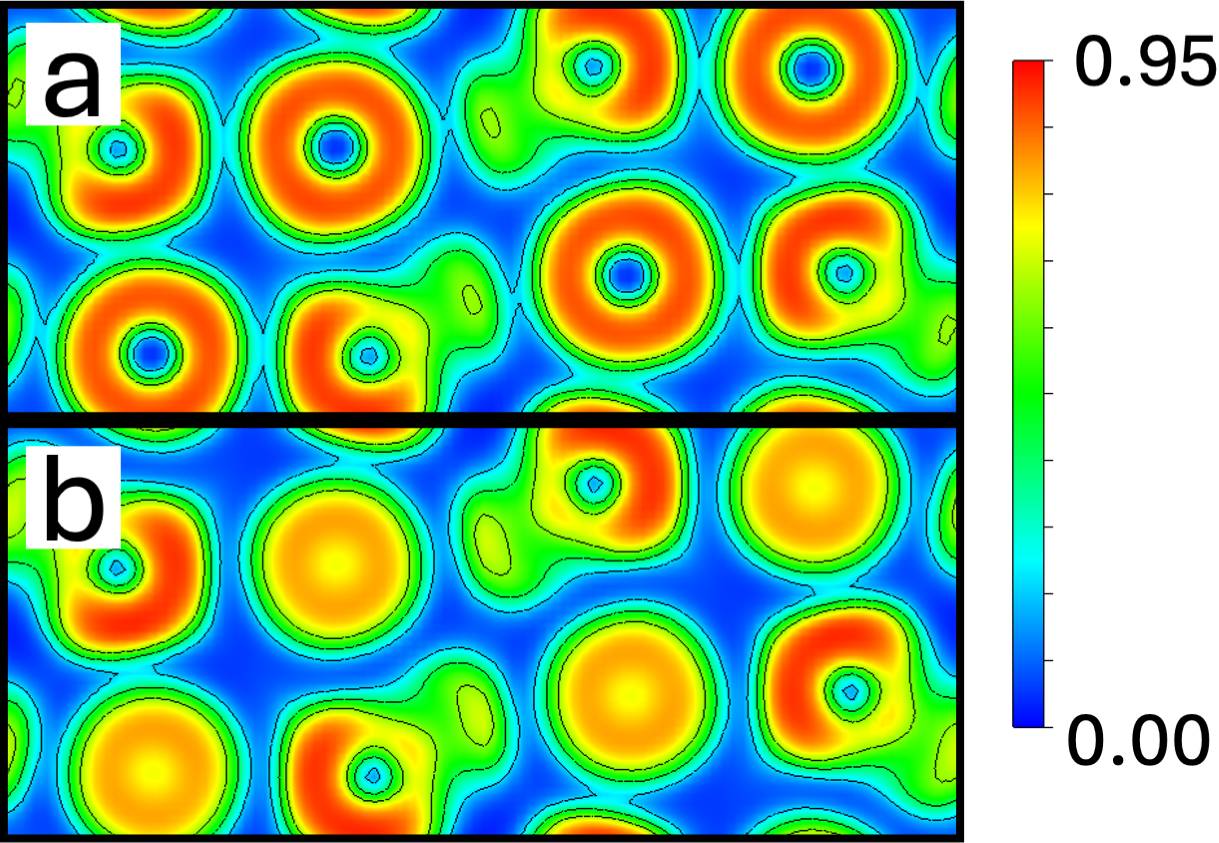}
    \caption{Electron localization function for \textbf{(a)} CeCN$_5$ at 90 GPa, and \textbf{(b)} TbCN$_5$ at 110 GPa. The plane cutting the unit cell is shown in Fig. \ref{fig:planes}. }
    \label{fig:elf_CeTb}
\end{figure}

\section{Magnetic structure of TbCN$_5$}
\begin{figure}[H]
    \centering
    \includegraphics[width=0.4\linewidth]{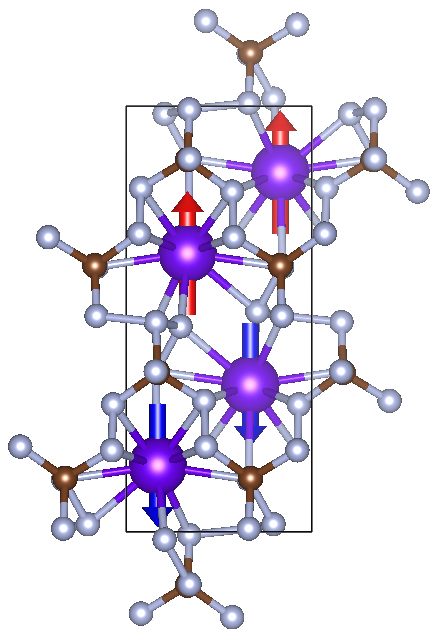}
    \caption{Magnetic configuration used in PBE+U calculations.}
    \label{fig:mag_conf_Tb}
\end{figure}

\section{Charge analysis}
Given a CHGCAR file containing the charge density sampled on a discrete grid, along with the atomic positions, the accumulated charge on each atom was estimated using a weighted Voronoi diagram–based partitioning scheme~\cite{masood2021visual, abrikosov2021topological}. For metal (M) atoms, ionic radii were used, specifically 0.97 Å for Ce$^{4+}$ and 1.04 Å for Tb$^{3+}$~\cite{shannon1976revised, database_ionic_radii}. For nitrogen and carbon atoms, van der Waals radii of 1.55 Å and 1.70 Å, respectively, were used.

Each grid point $p_i$ in the charge density grid was uniquely assigned to an atom in the unit cell according to the following criteria. A point $p_i$ was assigned to an M atom $a_j$ if $\mathrm{dist}(p_i, p(a_j)) < r(a_j)$, where $p(a_j)$ and $r(a_j)$ denote the position and radius of atom $a_j$, respectively, and $\mathrm{dist}(\cdot,\cdot)$ represents the Euclidean distance respecting the periodic boundary conditions. For all remaining grid points, assignment was performed with respect to nitrogen and carbon atoms by selecting the atom that minimizes the quantity $\mathrm{dist}(p_i, p(a_j)) - r(a_j)$. That is, $p_i$ was assigned to atom $a_j$ if
\[
\mathrm{dist}(p_i, p(a_j)) - r(a_j) < \mathrm{dist}(p_i, p(a_k)) - r(a_k), \quad \forall, k \neq j,
\]
where both $a_j$ and $a_k$ are either N or C atoms.

This procedure yields a partitioning of the grid into distinct regions corresponding to individual atoms within the unit cell. The charge associated with each atom was then obtained by integrating the charge density over its corresponding region. 
\begin{figure}
    \centering
    \includegraphics[width=0.3\linewidth]{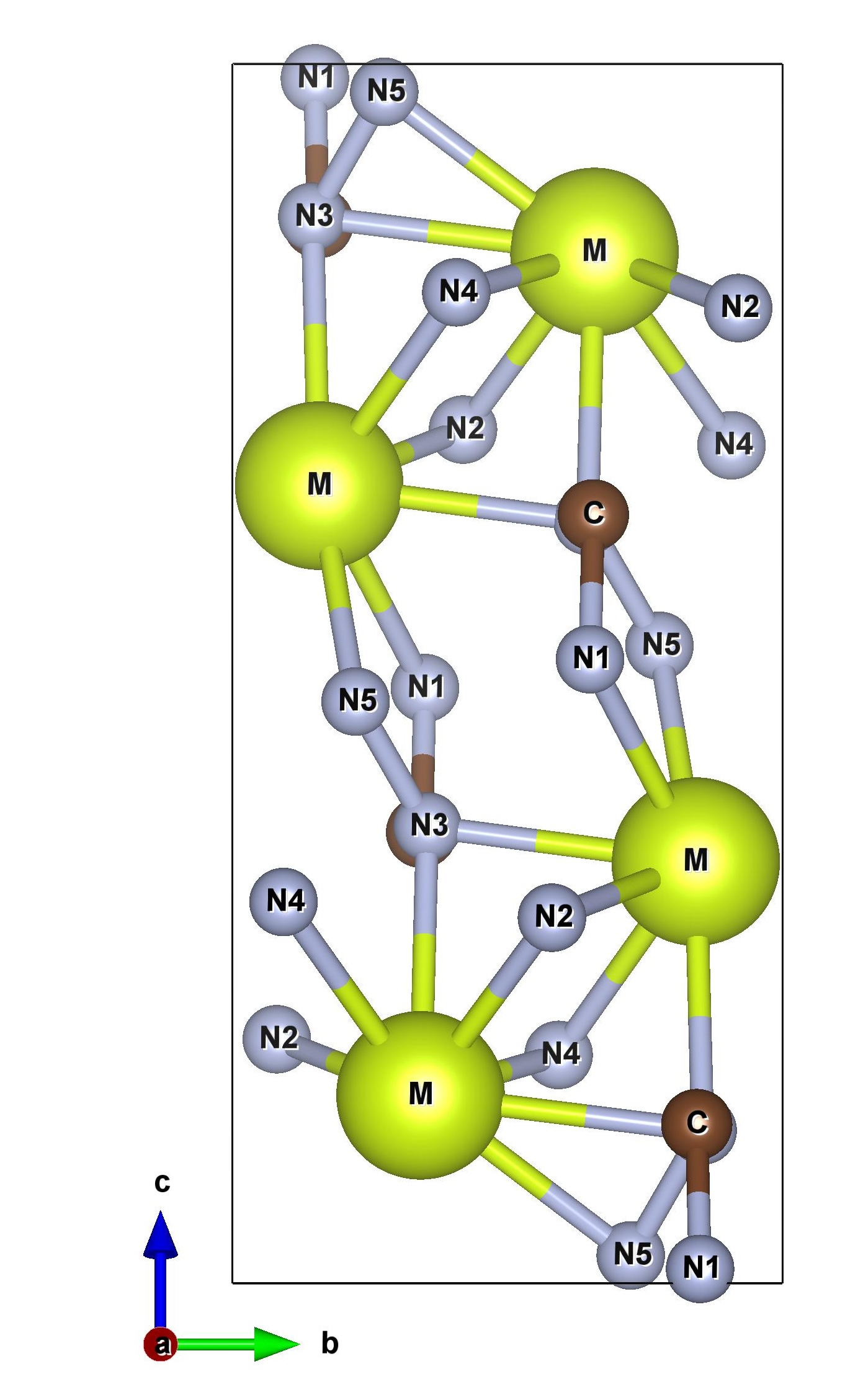}
    \caption{Explanation of atomic labeling for Fig. 5 in the main text.}
    \label{fig:placeholder}
\end{figure}

\bibliography{Ce_Tb_Manuscript.bib}